\documentclass[12pt, reqno]{article}

\usepackage{latexsym}
\usepackage{amsthm}
\usepackage{amsmath}
\usepackage{epsfig}
\usepackage{amssymb}
\usepackage{amsbsy}
\usepackage{color}
\usepackage{mathrsfs}

\newcommand{\be}{\begin{eqnarray}}
\newcommand{\ee}{\end{eqnarray}}

\definecolor{downstairs}{rgb}{0.09, 0.1328, 0.6388}
\definecolor{upstairs}{rgb}{0.698,0.1259,0.259}

\makeindex
\begin{document}

\baselineskip=18pt

\setcounter{footnote}{0}
\setcounter{figure}{0}
\setcounter{table}{0}

\begin{titlepage}

\begin{center}

{\Large \bf Fundamental BCJ Relation in ${\cal N}=4$ SYM From The Connected Formulation}

\vspace{0.5cm}

{\bf Freddy Cachazo}

\vspace{.1cm}

{\it Perimeter Institute for Theoretical Physics, Waterloo, Ontario N2J W29, CA}

\end{center}

\vspace{0.5cm}

\begin{abstract}

Tree-level amplitudes in ${\cal N}=4$ SYM can be decomposed into partial or color-ordered amplitudes. Identities relating various partial amplitudes have been known since the 80's. They are Kleiss-Kuijf (KK) identities. In 2008, Bern, Carrasco and Johansson (BCJ) introduced a new set of identities which reduce the number of independent partial amplitudes to $(n-3)!$. In recent years, several formulations for partial amplitudes have been discovered and shown to be equivalent to each other. These can be thought of as simple ``dualities" in the sense that different formulations make manifest different properties of the same object; the amplitude. One such formulation is the ACCK Grassmannian formulation which makes Yangian invariance of individual partial amplitudes manifest. A different formulation is the so-called connected formula introduced by Witten in twistor space and formulated in momentum space by Roiban, Spradlin and Volovich. It has been argued that the connected formula is ideal for studying properties which are related to the full amplitude, such as the KK relations, and not to particular partial amplitudes, like Yangian invariance. Following this logic, it is very natural to expect that the BCJ identities should have a very simple proof in the connected formulation. In this short note we show that this is indeed the case.

\end{abstract}

\bigskip
\bigskip

\end{titlepage}

\section{Tree Amplitudes and Dual Descriptions}

Tree level scattering amplitudes in ${\cal N}=4$ super Yang-Mills can be decomposed in terms of partial amplitudes,
\be
\label{full}
{\cal A}_n = \sum_{\sigma \in S_n/{\mathbb{Z}_n}}{\rm Tr}(T^{a_{\sigma (1)}}T^{a_{\sigma (2)}}\ldots T^{a_{\sigma (n)}}) A_n(\sigma (1),\sigma (2),\ldots , \sigma (n))
\ee
where the sum is over all permutations of $n$ labels modulo cyclic ones \cite{Dixon:1996wi}. Individual partial amplitudes or color-ordered amplitudes, e.g., $A_n(1,2,\ldots, n-1,n)$, are known to enjoy remarkable properties such as invariance under an infinite dimensional algebra known as the Yangian of $psu(2,2|4)$ whose level one generators are build using the particular ordering of the partial amplitude under consideration \cite{Beisert:2010jr}. The full amplitude, ${\cal A}_n$ is also known to satisfy interesting properties, some of which can be derived from the Lagrangian formulation, like the $U(1)$ decoupling identity or its generalizations known as the Kleiss-Kuijf (KK) relations \cite{Kleiss:1988ne}.

Luckily, two different but equivalent formations have been found where each class of identities becomes manifest. The formulation that makes manifest the Yangian invariance of $A_n(1,2,\ldots , n)$ is in terms of a contour integral on a Grassmmanian,
\begin{eqnarray}
\label{mome}
{\cal L}_{n,m} &=& \frac{1}{{\rm vol(GL}(m))}\int \frac{d^{m\times n}C}{(12\ldots m)(23\ldots m+1)\ldots (n1\ldots m-1)} \nonumber \\ & \times & \prod_{\alpha=1}^m \delta^4(C_{\alpha a}\tilde\eta_a) \delta^2(C_{\alpha a}\tilde\lambda_a) \int d^2\rho_\alpha \prod_{a=1}^n \delta^2(\rho_\beta C_{\beta a}-\lambda_a)
\end{eqnarray}
where $m$ is the $R$-charge of sector under consideration \cite{ArkaniHamed:2009dn, Mason:2009qx, ArkaniHamed:2009vw}. We use the convention that roman repeated indices in the arguments of delta functions are summer over the particle labels, e.g., the ``a" index in $C_{\alpha a}\tilde\eta_a$ indicates a sum over terms with $a\in \{ 1,2,\ldots ,n\}$.

The second formulation is the connected formula introduced by Witten \cite{Witten:2003nn} in twistor space and formulated and studied in momentum space by Roiban, Spradlin and Volovich \cite{Roiban:2004yf}. It turns out that the RSVW formula can be thought of as an integral over the Grassmannian $G(2,n)$ localized on maps of degree $m-2$ into $G(m,n)$ known as Veronese maps. The explicit form for $A_n(1,2,\ldots ,n)$ is given by
\be
\label{conn}
\frac{1}{{\rm vol(GL}(2))}\!\!\int\!\! \frac{d^{2n}\sigma}{(12)(23)\ldots (n1)}
\prod_{\alpha=1}^m \!\delta^{2}\! ( C_{\alpha a}[\sigma]\tilde\lambda_a )\delta^{4}( C_{\alpha a}[\sigma]\tilde\eta_a )\! \int\!\! d^2\rho_\alpha \prod_{b=1}^n\delta^2\! ( \rho_\beta C_{\beta b}[\sigma]\! -\! \lambda_b ) \ee
where the measure is $d^{2n}\sigma = d^2\sigma_1 d^2\sigma_2\ldots d^2\sigma_n$. The integrand is invariant under ${\rm GL(2)}$ transformations acting on each $\sigma_a = (\sigma_{(1)a},\sigma_{(2)a})$ as can easily be checked using that $(a,b)=\sigma_{(1)a}\sigma_{(2)b}-\sigma_{(2)a}\sigma_{(1)b}$ is ${\rm SL(2)}$ invariant and scales appropriately to cancel the transformation of the measure. The Veronese map is explicitly defined to be
$$ C_{\alpha a}[\sigma ] = \sigma_{(1)a}^{m-\alpha}\sigma_{(2)a}^{\alpha-1}.$$
In this formula each one-particle state is represented using an on-shell SUSY coherent state $|\lambda_a,\tilde\lambda_a, \tilde\eta_a \rangle$. Details on the connection between the $GL(2)$ invariant form used in this paper and the original RSVW formulation can be found in \cite{ArkaniHamed:2009dg}.

\section{Relations Among Partial Amplitudes}

In this note we concentrate on identities that relate different partial amplitudes. It is known that the independent set of identities are given by the KK relations and the fundamental Bern-Carrasco-Johansson (BCJ) identities \cite{Bern:2008qj} (for a review see \cite{Feng:2011gc}).

In order to illustrate why the connected formulation is useful to make manifest properties that relate different partial amplitudes manifest consider the $U(1)$ decoupling identity. This identity reads
\be
\label{uone}
\sum_{a=1}^{n} A(1,2,\ldots ,a,n+1,a+1,\ldots ,n) =  0.
\ee
We have chosen to express the identity in terms of $(n+1)$ partial amplitudes for later convenience.

The power of the connected formula (\ref{conn}) is that most elements in it are permutation invariant. In particular, all the information about the particular R-charge sector is encoded in the delta functions which are permutation invariant. The only part of (\ref{conn}) which knows about the color-ordering is the MHV-like factor $1/(12)(23)\ldots (n1)$. This means that if the identity (\ref{uone}) is supposed to be a property in all R-charge sectors, then it'd better be a property of the MHV-like factors or become something permutation invariant which vanishes on the support of the delta functions. The fact that the $U(1)$ decoupling identity and its generalizations are manifest in the connected formula was stressed in the original RSV work \cite{Roiban:2004yf}.

Using (\ref{conn}) for each of the amplitudes in the identity (\ref{uone}) one finds that the following should hold
\be
\label{utwo}
\sum_{a=1}^{n} \frac{1}{(12)(23)\ldots (a-1,a)(a,n+1)(n+1,a+1)(a+1,a+2)\ldots (n-1,n)(n,1)}=0.
\ee
Of course, this is completely equivalent to the $U(1)$ decoupling identity for MHV amplitudes. However, it is instructive to review the proof. Factoring out from each term $(a,a+1)/(a,n+1)(n+1,a+1)$ one finds
\be
\label{uti}
\frac{1}{(12)(23)\ldots (n-1,n)(n,1)}\sum_{a=1}^n\frac{(a,a+1)}{(a,n+1)(n+1,a+1)}
\ee
where the prefactor is precisely an MHV-like factor for $n$ particles. The fact that the sum vanishes is the well-known eikonal identity (note that in the sum $a+1$ is defined to be $1$ for $a=n$) \cite{Dixon:1996wi}.

\section{Fundamental BCJ Relation}

Consider the form of the fundamental BCJ relation which was used by Tye and Zhang to prove it in the MHV sector \cite{Tye:2010kg}\footnote{The formula below differs from the one presented in \cite{Tye:2010kg} by some trivial relabeling.}
\be
\sum_{a=1}^n \left(\sum_{b=1}^a s_{n+1,b}\right)A(1,2,\ldots ,a,n+1,a+1,\ldots ,n) = 0
\ee
A simple rearrangement of terms leads to another form which is somewhat analogous to the $U(1)$ decoupling identity
\be
\label{bcj}
\sum_{b=1}^n s_{n+1,b}\left(\sum_{a=b}^n A(1,2,\ldots ,a,n+1,a+1,\ldots ,n)\right) = 0.
\ee
Note that the term $b=1$ is exactly the $U(1)$ decoupling identity and thus vanishes.

For the reader's convenience, let us rewrite the connected prescription formula for partial amplitudes with $n+1$ particles, {\it i.e.}, the partial amplitude $A(1,2,\ldots ,n,n+1)$ is given by
\be
\label{conne}
\int\!\!\! \frac{[d^2\sigma_1 d^2\sigma_2\ldots d^2\sigma_{n+1}]}{(1~2)(2~3)\cdots (n+1~1)}
\prod_{\alpha=1}^m \delta^{2}\left( C_{\alpha a}[\sigma]\tilde\lambda_a \right)\delta^{4}\!\!\left( C_{\alpha a}[\sigma]\tilde\eta_a \right) \int d^2\rho_\alpha \prod_{b=1}^{n+1}\delta^2\left( \rho_\beta C_{\beta b}[\sigma]-\lambda_b \right) \ee
where $[d^2\sigma_1 d^2\sigma_2\ldots d^2\sigma_n]$ is the measure over $G(2,n)$, $\sigma_a = (\sigma_{(1)a},\sigma_{(2)a})$ and
$$ C_{\alpha a}[\sigma ] = \sigma_{(1)a}^{m-\alpha}\sigma_{(2)a}^{\alpha-1}$$

Using (\ref{conne}) to write every amplitude in (\ref{bcj}) as an integral over $G(2,n)$, the object we would like to study is the integrand
\be
\label{suma}
\sum_{b=1}^n s_{n+1,b}\sum_{a=b}^n \frac{1}{(1,2)(2,3)\cdots (a-1,a)(a,n+1)(n+1,a+1)(a+1,a+2)\cdots (n1)}
\ee
Recall that here the factors $(a,b)$ denote the two by two determinants in the $\sigma$ matrix while $s_{n+1,b}=\langle n+1,b\rangle [n+1,b]$.

Following the same steps as in the $U(1)$ decoupling identity, it is natural to extract a ``soft-like" factor associated with particle $n+1$ from each term in the sum. More explicitly, the soft-like factor is $(a,a+1)/(a,n+1)(n+1,a+1)$. This means that (\ref{suma}) becomes
\be
\frac{1}{(1,2)(2,3)\ldots (n-1,n)(n,1)}\times \sum_{b=1}^n s_{n+1,b}\sum_{a=b}^n \frac{(a,a+1)}{(a,n+1)(n+1,a+1)}.
\ee
Each sum over $a$ can easily be carried out using the eikonal identity
$$ \sum_{a=b}^n \frac{(a,a+1)}{(a,n+1)(n+1,a+1)} = \frac{(b,1)}{(b,n+1)(n+1,1)}$$
The integrand of object we would like to prove is zero then becomes
\be
\label{wawi}
\frac{1}{(1,2)(2,3)\ldots (n-1,n)(n,1)}\times\frac{1}{(n+1,1)}\times \sum_{b=1}^n s_{n+1,b}\frac{(b,1)}{(b,n+1)}
\ee
where we pulled out the factor of $(n+1,1)$ out of the sum for later convenience.

If the sum in this expression were to vanish on the support of the delta functions in (\ref{conne}), then the BCJ relation would follow.

The first step is to note that the factor $\langle n+1,b\rangle$ in $s_{n+1,b}$ can be written in terms of $\rho_\alpha$, $\sigma_{n+1}$ and $\sigma_b$ by using that on the support of the delta functions of (\ref{conne})
$$\lambda_a = \sum_{\alpha=1}^m \rho_\alpha C_{\alpha a}(\sigma).$$
It then follows that
\be
\label{dodo}
\langle n+1,b\rangle = \sum_{\alpha,\beta} \langle \rho_\alpha, \rho_\beta \rangle C_{\alpha b}C_{\beta, n+1}.
\ee
Therefore $\langle n+1,b\rangle$ is a polynomial of degree $m-1$ in $\sigma_b$, In other words, under $(\sigma_{(1)b},\sigma_{(2)b})\to (t\sigma_{(1)b},t\sigma_{(2)b})$, the polynomial scales by factor of $t^{m-1}$. Moreover, the polynomial vanishes when $\sigma_b$ becomes proportional to $\sigma_{n+1}$, therefore it must have $(n+1,b)$ as a factor. Explicitly,
$$ \langle n+1,b\rangle = (n+1,b)\times P_{m-2}(\sigma_b)$$
where $P_{m-2}$ is a polynomial of degree $m-2$. In what follows it will be very important to keep in mind the obvious fact that the coefficients of $P_{m-2}(\sigma_b)$ do not depend on $b$.

Using this result in the sum in (\ref{wawi}) we find
\be
\sum_{b=1}^n P_{m-2}(\sigma_b)\times [\tilde\lambda_b~ \tilde\lambda_{n+1}](b~1)
\ee
where the factor $(n+1,b)$ that was in the denominator canceled with the same factor in $\langle n+1,b\rangle$.

Finally, note that we can freely extend the sum to include the term $b=n+1$ as the corresponding term manifestly vanishes. Let us also denote by $P_{m-1}(\sigma_b)$ the degree $m-1$ polynomial $P_{m-2}(\sigma_b)\times (b~1)$.

This means that we have
\be
\sum_{b=1}^{n+1} P_{m-1}(\sigma_b)\times [\tilde\lambda_b ~ \tilde\lambda_{n+1}].
\ee
Now recall that the delta functions in (\ref{conne}) imply that
\be
\label{halo}
\sum_{a=1}^{n+1} \sigma_{(1)a}^{m-\alpha}\sigma_{(2)a}^{\alpha -1}\tilde\lambda_a
 = 0
\ee
for all $\alpha$ in $\{ 1,\ldots ,m\}$.

If we multiply (\ref{halo}) by arbitrary coefficients $h_\alpha$ and sum over $\alpha$, we find that any polynomial of degree $m-1$ in $\sigma_a$, say $P_{m-1}(\sigma_a)$ produces an $n$ vector which is orthogonal to the $2$-plane in $\mathbb{C}^{n+1}$ given by $\tilde\lambda_a$. More explicitly,
\be
\sum_{a=1}^{n+1} P_{m-1}(\sigma_a)\tilde\lambda_a = 0
\ee
and this completes the proof.

\section{Conclusions: The Identity of RSVW Residues}

We have shown that the connected formulation of scattering amplitudes makes all known properties among partial amplitudes simple consequences of the MHV-like structure at its core and the fact that all information about different $R$-charge sectors is completely permutation invariant.

In the RSVW connected formula, one has to solve several polynomial equations and the amplitude is computed as a sum of the integrand (time a jacobian) over all solutions. For example, for $n=6$ and $m=3$ one finds $4$ solutions while for $n=7$ and $m=3$ one finds $11$ solutions.

A very important observation is that our proof shows that each individual solution satisfies both the KK and BCJ relations. It is fascinating that each residue has all the same global properties as the full amplitude.

It is known that the BCJ identities are the backbone of the Kaway-Lewellen-Tye relations \cite{Kawai:1985xq} which connect ${\cal N}=8$ supergravity amplitudes to products of ${\cal N}=4$ Yang-Mills amplitude. Given that the connected formulation of Yang-Mills amplitudes makes the BCJ relations manifest, it would be interesting to explore its connection to the KLT formulation of gravity amplitudes.

\section*{Acknowledgments}

We thank N. Arkani-Hamed, B. Feng, Y. Geyer, Y. Gu, S. He, and S. Rajabi for useful discussions. This research was supported in part by the NSERC of Canada and MEDT of Ontario.


\begin{thebibliography}{10}

\bibitem{Dixon:1996wi}
  For a review see: L.~J.~Dixon,
  ``Calculating Scattering Amplitudes Efficiently,''
  In *Boulder 1995, QCD and beyond* 539-582
  [hep-ph/9601359].

\bibitem{Beisert:2010jr}
  See chapter V of: N.~Beisert, C.~Ahn, L.~F.~Alday, Z.~Bajnok, J.~M.~Drummond, L.~Freyhult, N.~Gromov and R.~A.~Janik {\it et al.},
  ``Review of AdS/CFT Integrability: An Overview,''
  Lett.\ Math.\ Phys.\  {\bf 99}, 3 (2012)
  [arXiv:1012.3982 [hep-th]].

\bibitem{Kleiss:1988ne}
  R.~Kleiss and H.~Kuijf,
  ``Multi - Gluon Cross-sections And Five Jet Production At Hadron Colliders,''
  Nucl.\ Phys.\ B {\bf 312}, 616 (1989).

\bibitem{ArkaniHamed:2009dn}
N.~Arkani-Hamed, F.~Cachazo, C.~Cheung, and J.~Kaplan, ``{A Duality for The
  S-Matrix},'' 2009, arXiv:0907.5418~[hep-th].

\bibitem{Mason:2009qx}
  L.~J.~Mason and D.~Skinner,
  ``Dual Superconformal Invariance, Momentum Twistors and Grassmannians,''
  JHEP {\bf 0911}, 045 (2009)
  [arXiv:0909.0250 [hep-th]].

\bibitem{ArkaniHamed:2009vw}
N.~Arkani-Hamed, F.~Cachazo, and C.~Cheung, ``{The Grassmannian Origin Of Dual
  Superconformal Invariance},'' 2009, arXiv:0909.0483~[hep-th].

\bibitem{Witten:2003nn}
E.~Witten, ``{Perturbative Gauge Theory as a String Theory in Twistor Space},''
  {\em Commun. Math. Phys.}, vol.~252, pp.~189--258, 2004.

\bibitem{Roiban:2004yf}
  R.~Roiban, M.~Spradlin and A.~Volovich,
  ``On the Tree Level S Matrix of Yang-Mills Theory,''
  Phys.\ Rev.\ D {\bf 70}, 026009 (2004)
  [hep-th/0403190].

\bibitem{ArkaniHamed:2009dg}
  N.~Arkani-Hamed, J.~Bourjaily, F.~Cachazo and J.~Trnka,
  ``Unification of Residues and Grassmannian Dualities,''
  JHEP {\bf 1101}, 049 (2011)
  [arXiv:0912.4912 [hep-th]].

\bibitem{Bern:2008qj}
Z.~Bern, J.~J.~M.~Carrasco and H.~Johansson,
  ``New Relations for Gauge-Theory Amplitudes,''
  Phys.\ Rev.\ D {\bf 78}, 085011 (2008)
  [arXiv:0805.3993 [hep-ph]].

\bibitem{Feng:2011gc}
  B.~Feng and M.~Luo,
  ``An Introduction to On-shell Recursion Relations,''
  arXiv:1111.5759 [hep-th].

\bibitem{Tye:2010kg}
  H.~Tye and Y.~Zhang,
  ``Comment on the Identities of the Gluon Tree Amplitudes,''
  Phys.\ Rev.\ D {\bf 82}, 087702 (2010)
  [arXiv:1007.0597 [hep-th]].

\bibitem{Kawai:1985xq}
  H.~Kawai, D.~C.~Lewellen and S.~H.~H.~Tye,
  ``A Relation Between Tree Amplitudes of Closed and Open Strings,''
  Nucl.\ Phys.\ B {\bf 269}, 1 (1986).

\end{thebibliography}

\end{document}